\documentclass[secnumarabic,tightenlines,eqsecnum,floats,prd,twocolumn,superscriptaddress,showpacs,preprintnumbers,amsmath,amssymb,APS,nofootinbib]{revtex4-1}
\usepackage{bm}
\usepackage{amsmath,amssymb,amsthm}
\usepackage{amsfonts}
\usepackage{amssymb}
\usepackage{graphicx}
\usepackage{color}
\usepackage{nicefrac}
\usepackage{braket} 

\newcommand{\be}{\begin{equation}}
\newcommand{\ee}{\end{equation}}
\newcommand{\bea}{\begin{eqnarray}}
\newcommand{\eea}{\end{eqnarray}}

\begin{document}
\title{{\bf Adiabatic regularization for spin-$\textbf{1}$  fields}} 

\author{F. Javier Mara\~{n}\'on-Gonz\'alez}
\affiliation{ {\footnotesize Departamento de F\'isica Te\'orica and IFIC, Universidad de Valencia-CSIC,
    Burjassot-46100, Valencia, Spain}}

\author{Jos\'e Navarro-Salas}\email{jnavarro@ific.uv.es}
\affiliation{ {\footnotesize Departamento de F\'isica Te\'orica and IFIC,  Universidad de Valencia-CSIC,
    Burjassot-46100, Valencia, Spain}}

\date{October 18, 2023}

\begin{abstract}

We analyze the adiabatic regularization scheme to renormalize Proca fields in a four-dimensional Friedmann-Lema\^{i}tre-Robertson-Walker spacetime. The adiabatic method is well established for scalar and spin-$1/2$ fields, but is not yet  fully understood  for spin-$1$ fields. 
We give the details of the construction and show that, in the massless limit, the renormalized stress-energy tensor of the Proca field is closely related to that of a  minimally coupled scalar field.  Our result is in full agreement with other approaches, based on the effective action, which also show a discontinuity in the massless limit. 
The scalar field can be naturally regarded as a St\"ueckelberg-type field. We also test the consistency of our results in de Sitter space.

\end{abstract}


\maketitle

\section{Introduction}

In a Friedmann-Lema\^{i}tre-Robertson-Walker (FLRW) spacetime it is particularly useful to  implement  renormalization using the  adiabatic regularization method. It was systematically introduced in \cite{parker-fulling} to renormalize the stress-energy tensor of massive scalar fields (see \cite{birrell-davies, fulling, parker-toms, hu-verdaguer} for textbook introductions).  The method is based on the WKB-type adiabatic expansion of the field modes, for which  the all-orders adiabatic condition is equivalent to the Hadamard condition for the two-point function \cite{pirk}. The conformally invariant case can be obtained by setting the mass to zero 
 at the end of the calculations \cite{Hu} (for the hyperbolic FLRW case see  \cite{bunch80} and  for the closed FLRW universes \cite{Anderson-Parker}). 
One  of the advantages of the adiabatic method is its efficiency   in terms of numerical implementation \cite{Anderson}. More recently the adiabatic regularization has been extended to spin-$1/2$ fields \cite{LNT, RNT, Ghosh:2016, BFNV, BNP}. The adiabatic expansion then differs 
significantly  from the WKB-type expansion of the spin-$0$ field. The conformal anomaly can also be reproduced, in the limit $m\to 0$, from the  fourth order in the adiabatic expansion of the stress-energy tensor.

The renormalization of a massive spin-$1$ field in a curved spacetime is often considered within the framework of St\"ueckelberg formalism \cite{janssen-dullemond, frob-higuchi, belokogne-folacci}, which includes a minimally coupled scalar field in addition  to the spin-$1$ field $A_\mu$. The covariant quantization of St\"ueckelberg's massive electromagnetism  also involves a gauge breaking term and ghost fields. The main reason to proceed this way is that, in St\"ueckelberg's theory, all  relevant two-point functions admit a Hadamard representation. In contrast, working in the conventional Proca description, the  constraint $\nabla_\mu A^\mu=0$ causes the Feynman propagator associated with the vector field $A_\mu$ not to admit a Hadamard representation.

Although  adiabatic regularization can also be applied to St\"ueckelberg's electromagnetism  or a gauge-fixed theory \cite{chimento-cossarini, agullo, chu-koyama, ordines-carlson}, it is closer to the original idea of the adiabatic method to approach the problem following the  canonical  quantization  route. {\color{black} The implementation of the adiabatic method requires a field theory with a nonvanishing initial mass. Therefore, in general,  conformally invariant theories are analyzed in the adiabatic regularization  by taking the massless limit at the end of the calculations. This makes the application of the method  for a  gauge field with zero mass more delicate, due to the discontinuity in the number of physical degrees of freedom.}
The systematic study of the adiabatic regularization of the Proca field is the main objective of this paper. 
We will analyze  in detail the behavior of the renormalized theory in the massless limit. 
We will focus on the required renormalization subtractions for the  two-point function  at coincidence $\langle A_\mu A^\mu \rangle$ and the stress-energy tensor $\langle T_{\mu\nu}\rangle$.
{\color{black} The main result of our analysis can be summarized in the following expression 
\be \label{main1}\langle T^{\text{Proca}}_{\mu\nu} \rangle -\langle T^{\text{Scalar }\xi=0}_{\mu\nu}\rangle \sim _{m\to 0}  \langle T^{\text{Maxwell}}_{\mu\nu} \rangle   \ , \ee
where $\langle T^{\text{Scalar }\xi=0}_{\mu\nu}\rangle$ refers to the vacuum expectation values of the stress-energy tensor of a minimally coupled scalar field. This auxiliary scalar field can be  naturally interpreted as a St\"ueckelberg  field.
}
Our results are fully consistent  with  those obtained from the one-loop effective action 
\cite{buchbinder-shapiro21, buchbinder-berredo-shapiro},  and also with  the quantization of the massless and minimally coupled scalar field in de Sitter space \cite{allen-folacci}.


The paper is organized as follows. In Sec. \ref{scalars} we provide a general introduction to the adiabatic method for scalar fields. 
Section \ref{proca} is devoted to a systematic analysis of the massive Proca field  and the corresponding WKB-type ansatz for the transverse and longitudinal polarizations. In Sec. \ref{regularization} we focus on the evaluation of the adiabatic subtraction terms for $\langle T_\alpha^\alpha \rangle  = - m^2  g_{\alpha\beta} \langle A^\alpha A^\beta \rangle$ and study the massless limit. 
In Sec. \ref{comparison} we show that our results are in agreement  with those obtained from other approaches. 
Finally, we summarize our conclusions in Sec. \ref{conclusions}. Throughout this paper, we use units in which $\hbar = c = 1$. Our  conventions for the signature of the metric and the curvature tensor follow Refs. \cite{birrell-davies, parker-toms}.

\section{Adiabatic regularization for scalar fields}\label{scalars}

To give an overview of the main ideas of the adiabatic method let us consider a scalar field $\phi$ on a spatially flat FLRW background 
\be ds^2 = dt^2 - a^2(t) (dx^2 + dy^2 + dz^2) \ , \ee
satisfying the field equation
\be (\Box + m^2 + \xi R) \phi =0 \ . \ee
Canonical quantization proceeds by expanding the field  in a complete orthonormal set of elementary solutions
\be \label{modedec}\phi (t,\vec{x}) = \int \mathrm{d}^3 k \, \left(a_{\vec k} f_{\vec k} (t,\vec{x}) + a_{\vec k}^{\dagger}f^*_{\vec k} (t,\vec{x})\right) \ . \ee
It is useful to  write the field modes as 
\be f_{\vec k}(t,\vec{x}) = \frac{1}{\sqrt{2(2\pi)^3 a^3(t)}} h_k(t) e^{i\vec k\cdot\vec x} \ , \ee
where $h_k(t)$ obeys the equation
\be \label{waveq}\ddot h_k + (\omega_k^2  + \sigma) h_k =0 \ . \ee
Here $\omega_k^2 = k^2/a^2 + m^2$ and $\sigma$ is defined by 
\be \sigma = (6\xi -3/4) \dot a^2/a^2 +(6\xi -3/2)\ddot a/a \ . \ee  The field modes have been assumed to be orthonormal with respect to the Klein-Gordon inner product
\be (\phi_1, \phi_2) = i\int d^3 x a^3 (\phi_1^* \partial_t \phi_2 - \partial_t \phi_1^* \phi_2) . \ee
Hence, the functions $h_k$ must satisfy the normalization condition
\be \label{wronskian} h_k \dot h_k^* - h_k^* \dot h_k = 2i \ . \ee
These functions are also constrained to behave, for large $k$, as 
\be \label{WKB} h_k(t) \sim \frac{1}{\sqrt{W_k(t)}} e^{-i\int^t W_k(t') \mathrm{d}t'} \ , \ee
where the leading term for $W_k(t)$ is the physical frequency $\omega_k$. Hence for large $k$ the modes behave as the Minkowskian modes. The higher-order terms 
\be \label{W} W_k(t) \sim \omega_k +  {\color{black} W_k^{(2)}} + \cdots  \ee
are systematically obtained 
by  plugging the WKB-type ansatz (\ref{WKB}) into Eq. (\ref{waveq}).  The leading term  determines univocally the higher-order terms in the adiabatic expansion. One gets
\be 
W_k^{(2)}= \frac{1}{2} \omega_k^{-1/2} \frac{d^2}{dt^2} \omega_k^{-1/2} + \frac{1}{2} \omega_k^{-1} \sigma\ ,
\ee
where all the terms of odd adiabatic order are zero.

The above adiabatic expansion serves to identify the ultraviolet divergences that plague the vacuum expectation values quadratic in fields, as  the stress-energy tensor $\langle T_{\mu\nu} \rangle$ or the two-point function  at coincidence $\langle \phi^2\rangle$. The quantum states are assumed to obey the (asymptotic) adiabatic condition (\ref{WKB}). 
To illustrate how the adiabatic method works let us consider the formal expression for the two-point function
\be \label{phi2} \langle \phi^2(t,\vec{x})\rangle = \frac{1}{4\pi^2 a^3(t)}\int_0^\infty \mathrm{d}k\, k^2  |h_k(t)|^2 \ . \ee
Following \cite{parker-toms} one can perform the asymptotic expansion
\be \label{expansionphi2}\langle \phi^2\rangle \sim  \frac{1}{4\pi^2 a^3}\int_0^\infty \!\!\!\!\! \mathrm{d}k\, k^2  \ {\color{black} W_k^{-1}} \ , \ee 
where
\be W_k^{-1} = \omega_k^{-1} + (W_k^{-1})^{(2)} + (W_k^{-1})^{(4)} +\cdots   \ee
is the corresponding adiabatic expansion of $W_k^{-1}$ according to (\ref{W}). The expansion (\ref{expansionphi2}) captures the ultraviolet  divergences in the $k$-integral for the two-point function in (\ref{phi2}). These divergences are independent of the  specific vacuum state [among the adiabatic vacua obeying the asymptotic condition (\ref{WKB})] and should be removed by renormalization. The required renormalization subtractions are then obtained from (\ref{expansionphi2}).  We immediately observe that the divergences  appear in the zero and second adiabatic orders, while the terms of fourth adiabatic order and beyond   are all finite. Therefore, one should only subtract the full zero and second adiabatic order terms. The renormalized expression for $\langle \phi^2\rangle_{\rm ren}$ is then constructed by the momentum integral
\be \label{phi2ren} \langle \phi^2\rangle_{\rm ren} = \frac{1}{4\pi^2 a^3}\int_0^\infty \!\!\!\! \mathrm{d}k\, k^2 \left( |h_k|^2 -   \omega_k^{-1} - (W_k^{-1})^{(2)}  \right) \ . \ee
The renormalization of $\langle T^{\mu\nu} \rangle$ requires subtracting until and including the fourth adiabatic order.

For our purposes it is useful to consider the trace of the stress-energy tensor $\langle T^\mu_\mu\rangle$. The operator expression of the trace, once the equations of motion have been used, is
\be \label{truetrace}T^\alpha_\alpha = (6 \xi - 1) \partial^\alpha \phi \partial_\alpha \phi + \xi (1- 6 \xi)R \phi^2 + (2-6 \xi) m^2 \phi^2 \ .  \ee
For $\xi=1/6$ the vacuum expectation values of the trace reduces to
\be \langle T^\alpha_\alpha \rangle = m^2 \langle\phi^2 \rangle \ . \ee
The adiabatic subtraction terms to fourth order are finite in the limit $m^2\to 0$ and they define the conventional trace anomaly of a massless and conformally coupled scalar field
\begin{equation}
\begin{split}
    \langle T^\alpha_\alpha \rangle &= \lim_{m\to 0} -m^2 \langle\phi^2 \rangle^{(4)} \\
&= \frac{1}{2880\pi^2} \left[-\left(R_{\mu\nu}R^{\mu\nu} -\frac{1}{3}R^2\right) + \Box R\right]\ .
\end{split}
\end{equation} 
For arbitrary $\xi$, the above expression contains  additional terms
\begin{equation}
\begin{split}    
\lim_{m\to 0} -m^2 \langle\phi^2 \rangle^{(4)} &= \frac{1}{2880\pi^2} \left[-\left(R_{\mu\nu}R^{\mu\nu} -\frac{1}{3}R^2\right) \right. \\
&+ \left. \left(\vphantom{\frac13}6-30\xi\right)\Box R-90\left(\xi-\frac{1}{6}\right)^2 R^2\right] \ ,
\end{split}
\end{equation}
often referred to as the anomalous contribution to the trace \cite{birrell-davies}. The above expression is equal to $-a_2(x)/16\pi^2$, where $a_2(x)$ is the second coefficient arising in the DeWitt-Schwinger  proper-time formalism.

Going back to (\ref{truetrace}) we see that there is  a nonanomalous contribution from kinetic and mass terms for general $\xi$, in contrast with the conformal scalar, $\xi=1/6$, where only the mass term contributes. Taking the vacuum expectation values we get 
\bea \langle T^\alpha_\alpha \rangle &=& \int d^3 k [(6 \xi - 1) \partial^\alpha f_{\vec k} \partial_\alpha f^*_{\vec k} +
 \xi (1- 6 \xi)R |f_{\vec k}|^2 \nonumber \\ &+& (2-6 \xi) m^2 |f_{\vec k}|^2 ]. \eea
Plugging now the adiabatic expansion  for the modes one can obtain the general expression for the adiabatic expansion of $\langle T^\alpha_\alpha \rangle$. For future use, we are particularly  interested in the minimally coupled case, $\xi=0$, and the contribution up to  the fourth adiabatic order.  One gets
\bea \label{adbBunch}
    &&\langle T^\alpha_\alpha \rangle^{(0-4)}_{\xi=0} = -\frac{5 m^2 \dot a^2}{48 \pi^2 a^2}  + \frac{m^2 \ddot a}{48 \pi^2 a} - \frac{3 \dot a^4}{32 \pi^2 a^4} 
    + \frac{233 \dot a^2 \ddot a}{480 \pi^2 a^3} \nonumber \\ &+& \frac{17 \ddot a^2}{240 \pi^2 a^2} + \frac{9 \dot a \dddot a}{160 \pi^2 a^2} 
    - \frac{11 \ddddot a}{480 \pi^2 a} \nonumber \\
    &+&\frac{1}{(2\pi)^2 a^3}\int_0^\infty  dk \ |\vec k|^2\left[ \frac{m^2}{\omega_k} + \frac{m^2 \dot a ^2}{2 a^2 \omega_k^3} + \frac{m^2 \ddot a}{a \omega_k^3} \right. \nonumber \\
    &+& \left.  \frac{\ddot a}{a \omega_k} + \frac{5 \dot a^2 \ddot a}{4 a^3 \omega_k^3} 
    - \frac{\ddot a^2}{4 a^2 \omega_k^3} - \frac{3 \dot a \dddot a}{4 a^2 \omega_k^3} - \frac{\ddddot a}{4 a \omega_k^3} \right]\ . \ 
\eea
We observe immediately that there are ultraviolet divergent terms, even at fourth adiabatic order, that do not cancel in the massless limit. Those terms are removed in the case $\xi=1/6$ and the finite terms take an explicit covariant form only when $\xi=1/6$. We will see how, for the minimally coupled case, those terms acquire a clear meaning when they are combined with similar expressions for the Proca field.

\section{Adiabatic expansion for the Proca field} \label{proca}

The action for the Proca {\color{black}fields} in curved space is
\be \label{Sproca} S= \int d^4x \sqrt{-g}\left(-\frac{1}{4}F_{\mu\nu}F^{\mu\nu} +\frac{1}{2}m^2 A^\mu A_\mu\right) \ , \ee
which can also be rewritten, up to surface terms, as
\be \label{Sproca2} S= \frac{1}{2}\int d^4x \sqrt{-g}A_\nu\left( g^{\mu\nu} \Box - \nabla^\nu\nabla^\mu + R^{\mu\nu} +m^2 g^{\mu \nu}\right) A_\mu  \ . \ee
The field equations derived from (\ref{Sproca}) are
\be \label{feqsP}\nabla_\mu F^{\mu\nu} + m^2 A^\nu =0 \ . \ee
Taking a covariant divergence one gets
\be \label{fep}\nabla_\nu \nabla_\mu F^{\mu\nu} + m^2 \nabla_\nu A^\nu =0 \ . \ee
The properties of the commutator $[ \nabla_\nu, \nabla_\mu]$ in terms of the Riemann curvature ensure the identity $\nabla_\nu \nabla_\mu F^{\mu\nu}=0$, thus enforcing the scalar constraint
\be \label{subc}\nabla_\nu A^\nu=0 \ . \ee 
Before restricting the theory to a FLRW spacetime it is convenient to recall the canonical quantization of the theory in Minkowski space

\subsection{The Proca field in Minkowski space}

The conserved inner product for the Proca field is defined as ($A_1$ and $A_2$ are two complex solutions of the Proca field equations and $\eta_{\mu\nu}$ is the Minkowski metric)
\be (A_1, A_2) = -i \int d^3x \eta_{\mu\nu}(A_{1}^{*\mu} \partial_0 A_{2 }^\nu - \partial_0 A_{1}^{*\mu}  A_{2 }^\nu) \ . \ee 
A complete set of orthonormalized modes is given by 
\be A^{\mu (r)}_{\vec k} = \epsilon^{\mu  (r)}(\vec k) \frac{e^{-ikx}}{\sqrt{2(2\pi)^3 \omega_k}} \ , \ee
where $kx= \omega_k t - \vec k \vec x$ and $x^\mu= (t, x^1, x^2, x^3) \equiv (t, \vec x)$,  $k^\mu = (\omega_k, k^1, k^2, k^3)= (\omega_k, \vec k)$.  Three independent (in general complex) polarization  vectors can be chosen as 
\be \epsilon^{\mu (1)}(\vec k)= (0, \vec \epsilon^{\ 1}(\vec k))\ , \ee
\be \epsilon^{\mu (2)}(\vec k)= (0, \vec \epsilon^{\ 2}(\vec k))\ , \ee
\be \label{lp0}\epsilon^{\mu (3)}(\vec k)= (\frac{|\vec k|}{m}, \frac{\omega_k}{m|\vec k|}\vec k) \ , \ee
with $|\vec \epsilon^{\ 1}|=1=|\vec \epsilon^{\ 2}|$, $\vec \epsilon^{\ 1}\cdot  \vec \epsilon^{\ 2}= 0$, and $\vec k \cdot \vec \epsilon^{\ 1}= 0= \vec k \cdot \vec \epsilon^{\ 2}$.  

The above set of polarization vectors, depending on the reference vector $\vec k$,  obey
\be k_{\mu}\epsilon^{\mu  (r)}(\vec k)=0 \ee
\be \epsilon^{\mu  (r)*}\epsilon_{\mu}^{\ (s)} = -\delta^{rs} \ . \ee
One can easily check that the positive-frequency modes are orthonormal with the above inner product
\be (A^{(r)}_{\vec k}, A^{(s)}_{\vec k'}) = \delta^{3}(\vec k -\vec k')\delta^{rs} \ee
As a consequence, the creation and annihilation operators in the expansion
\be A_\mu = \sum_{r=1}^3\int \frac{d^3k}{\sqrt{2(2\pi)^3 \omega_k}}\left(a^r_{\vec k} \epsilon_\mu^{(r)}(\vec k)e^{-ikx} + a^{r\dagger}_{\vec k} \epsilon_\mu^{(r)*}(\vec k)e^{ikx}\right)\ee
obey the usual commutation relations
\be [a^r_{\vec k}, a^{s\dagger}_{\vec k'}]= \delta^{rs}\delta^3(\vec k -\vec k')\ee

\subsection{The Proca field in a FLRW spacetime}

As for the scalar field, let us assume a spacetime metric of the form
\be ds^2 = dt^2 -a^2(t) d\vec x^2 \ . \ee
The only independent nonzero Christoffel symbols are
\be \Gamma^{0}_{ij}= \delta_{ij} \dot a a \ , \ \ \ \  \Gamma^i_{0j}= \delta_{ij}\frac{\dot a}{a} \ . \ee
The field equations (\ref{feqsP}) can be reexpressed as 
\be g^{\mu\rho}(\partial_\rho F_{\mu\nu}-\Gamma^\gamma_{\rho\mu}F_{\gamma\nu}-\Gamma^{\gamma}_{\rho\nu}F_{\mu\gamma}) + m^2 A_\nu =0 \ , \ee
and using the above metric and Christoffel symbols one gets
\be \label{feqsP2} \partial^\mu \partial_\mu A_\nu - \partial^\mu \partial_\nu A_\mu + \frac{\dot a}{a}(\partial_0 A_\nu - \partial_\nu A_0) + m^2 A_\nu =0 \ . \ee

Furthermore, the inner product for the Proca field in the FLRW spacetime takes the form 
\bea \label{ipPcs} (A_1, A_2) &=&  -i \int d^3x a^3 g_{\mu\nu}[A_{1}^{*\mu} \nabla_0 A_{2 }^\nu - (\nabla_0 A_{1}^{*\mu})  A_{2 }^\nu] \nonumber \\
&=& -i \int d^3x a^3 \{[A_{1}^{*0} \nabla_0 A_{2 }^0 - (\nabla_0 A_{1}^{*0})  A_{2 }^0] \nonumber \\
&-& \sum_ja^2 [A_{1}^{*j} \nabla_0 A_{2 }^j - (\nabla_0 A_{1}^{*j})  A_{2 }^j] \}\ . \eea
Taking into account that 
$ \nabla_0A^0 = \partial_0 A^0$, $\nabla_0 A^j =  \partial_0 A^j +\frac{\dot a}{a}A^j$,
the inner product can be expressed as 
\bea \label{ipPcs2} (A_1, A_2) &=&  
 -i \int d^3x a^3 \{[A_{1}^{*0} \partial_0 A_{2 }^0 - (\partial_0 A_{1}^{*0})  A_{2 }^0] \nonumber \\
 &-& \sum_ja^2 [A_{1}^{*j} \partial_0 A_{2 }^j - (\partial_0 A_{1}^{*j})  A_{2 }^j] \}\ . \eea

The explicit form of the inner product suggests a natural  ansatz for the modes of the Proca field. It is convenient to distinguish between the transverse  modes (i.e., $r=1,2$) and the longitudinal modes $r=3$.

\subsubsection{Transverse modes} A convenient way to write the  transverse modes is
\be \label{tp}A^{\mu (r)}_{\vec k} = \epsilon^{\mu  (r)}(\vec k) \frac{e^{i\vec k \vec x}}{a\sqrt{2(2\pi a)^3 }} h_k(t) \ , \ \ \ \ \ \ \ r=1,2 \ . \ee
Note that the constraint $\nabla_\mu A^{\mu (r)}=0$ is trivially satisfied for $r=1,2$. 
The normalization condition for the  transverse modes 
\be (A^{(r)}_{\vec k}, A^{(s)}_{\vec k'}) = \delta^{rs}\delta^3(\vec k - \vec k') \ , \ee
becomes
\be \label{Wc}   \dot h_k^{*} h_k - h_k^{*}\dot h_k= 2i \ . \ee

The differential equation for $h_k(t)$ is obtained by plugging (\ref{tp}) in (\ref{feqsP2}). One gets a similar equation to (\ref{waveq})
\be  \label{heq} \ddot h_k + \left[\omega_k^2 + \left(6\xi -\frac34\right) \left(\frac{\dot a}{a}\right)^2 +\left(6\xi - \frac32\right) \frac{\ddot a}{a} \right]h_k =0 , \ee
with $\xi=1/6$. As for the scalar field, the normalization  (\ref{Wc})  is the Wronskian condition for   the above equation.
The transverse polarization modes behave as two scalar fields of mass $m$ with conformal coupling. In the massless limit they are natural candidates to describe 
two conformally invariant degrees of freedom with a conformal anomaly matching exactly the conformal anomaly of two scalar fields with $\xi=1/6$.  This will be confirmed by the analysis of the renormalized stress-energy tensor. 
\subsubsection{Longitudinal modes}

The   ansatz for the longitudinal modes  is more involved since the constraint $\nabla_\mu A^\mu=0$ is  nontrivial for $r=3$.  In fact, if $A^{0} \neq 0$ it takes the form
\be \partial_0 A^0 + 3\frac{\dot a}{a} A^0 + \partial_i A^i =0 \ . \ee
The ansatz is 
\be \label{Chimento_3ansatz} A^{\mu (3)}_{\vec k} =   \frac{\omega_k}{mk} \left(\frac{|\vec k|^2 {\color{black}\mathcal{W}}}{\omega_k^2}, \vec k\right) \frac{e^{i\vec k \vec x}}{a\sqrt{2(2\pi a)^3 }} l_k(t).  \ee

\subsubsection{Normalization}

The following ansatz
\be A^\mu = \epsilon^\mu(\vec k,t) \frac{e^{i \vec{k}\vec{x}}}{a \sqrt{2 (2 \pi a)^3}} f_k(t), \ee
works for both transverse [$\epsilon^\mu \to (0, \vec \epsilon^{\,(1,2)}(k))$, $f_k \to h_k$] and longitudinal [$\epsilon^\mu \to \epsilon^{\mu (3)}(\vec k, t)$, $f_k \to l_k$]. 

Imposing the appropriate normalization with respect to the inner product produces the following relation: 
\be \label{longitudinal_normaliz} \mathfrak{Im}\left\{ \epsilon_\mu^* \dot \epsilon^\mu |f_k|^2 + \epsilon^*_\mu \epsilon^\mu f_k^*\left( \dot f_k - \frac{5\dot a f_k}{2 a} \right) + \epsilon_j^* \epsilon^j |f_k|^2 \frac{\dot a}{a}\right\} = a^2.\ee
For the transverse case this simplifies considerably, since 
\be \label{eps_norm} \epsilon^{(r)*}_\mu \epsilon^{\mu (r)} = - a^2, \;\;\; r=1,2 \ee 
and (\ref{Wc}) is recovered. Equation (\ref{eps_norm}) follows from $\epsilon^{\mu(1,2)} = (0, \vec \epsilon^{\,(1,2)}(k))$ and  $\vec{\epsilon}^{\,(1,2)*} \cdot \vec{\epsilon}^{\,(1,2)} = 1$. 


For the longitudinal case, we can also be tempted to  choose (\ref{eps_norm}). However, this turns out to be inconsistent and we have to follow an alternative route. 
Using the  definition for the longitudinal polarization implicit in (\ref{Chimento_3ansatz}) [$\epsilon^{\mu (3)}= \frac{\omega_k}{mk} (\frac{|\vec k|^2 \mathcal{W}}{\omega_k^2}, \vec k)$]
we get 
\be \dot \epsilon^{\mu(3)} = \left( \frac{|\vec k|^3 \dot a {\color{black} \mathcal{W}}}{a^3 m \omega_k^3} + \frac{k {\color{black} \mathcal{\dot W}}}{m \omega_k}, -\frac{k \dot a}{m\omega_k a^3}\vec{k} \right), \ee
and 
\be  \label{eps_norm_modified} \epsilon^{(3)*}_\mu \epsilon^{\mu (3)} = - a^2 g_{\mathcal {W}}(t), \;\;\; g_{\mathcal{W}}(t) = \left(\frac{1}{\omega_k^2}-\frac{1}{m^2}\right)|\mathcal{W}|^2\ + \frac{\omega_k^2}{m^2},   \ee
with $g_{\mathcal{W}}(t)$ a scalar function. 
Hence, the normalization condition becomes
\be \mathfrak{Im}\left\{\frac{|\vec k|^2}{a^2 m^2 \omega_k^2} |l_k|^2 {\mathcal {W}}^* \mathcal{\dot W} - g_{\mathcal{W}} l_k^* \dot l_k \right\} = 1 \ . \ee 
Using the two independent  conditions of the Proca constraint $\nabla_\mu A^{\mu (3)}$, namely
\be \label{nabla_3constraint} 0=\left( \frac{|\vec k|^2 \dot a}{\omega_k^2 a^3} +\frac{\dot l_k}{l_k} + \frac{ \dot a}{2 a} \right) \mathcal{W} + \mathcal{\dot W}  + i \omega_k^2\ ,  \ee
we constrain $\mathcal W$ by
\be \frac{|\vec k|^2 |l_k|^2}{a^2 m^2} \mathfrak{Re}\mathcal{W} + \frac{\omega_k^2}{m^2} \mathfrak{Im}\{l_k^* \dot l_k\} = -1,\ee
or in a simpler way
\be \label{normalizationReW} \mathfrak{Re}\mathcal{W} = \frac{\left(m^2 + \omega_k^2 \mathfrak{Im}\{l_k^* \dot l_k\}\right)}{|l_k|^2\left(m^2-\omega_k^2\right)}. \ee

\subsubsection{Longitudinal equation of motion}

To obtain the motion equations for the longitudinal polarization, we need to substitute the ansatz (\ref{Chimento_3ansatz}) into (\ref{feqsP2}). The extra constraint $\nabla_\mu A^{\mu (3)}=0$ emerges from the equations of motion, so we can ignore it for now. From the $\nu=0$ component we get
\be A_0 = -\frac{i}{a^2 \omega_k^2} \sum_i k^i \dot A_i, \ee 
which removes $\mathcal W$ from the spatial equations of motion but also acts as a constraint, from which we get
\be \mathcal{W} = i \left( \frac{\dot l_k}{l_k} - \frac{3\dot a}{2a} +\frac{m^2 \dot a}{\omega_k^2 a} \right) \ee
or equivalently
\be \label{nuzero_longit} \mathfrak{Re}\mathcal{W} = - \frac{\mathfrak{Im}\{l_k^* \dot l_k\}}{|l_k|^2}, \;\;\; \mathfrak{Im}\mathcal{W} = \frac{\mathfrak{Re}\{l_k^* \dot l_k\}}{|l_k|^2} - \frac{3\dot a }{2a} + \frac{m^2 \dot a }{\omega_k^2 a}.  \ee
From the above one can derive the corresponding equation of motion for $l_k$
\begin{equation} \label{leq}
\begin{split}    
\ddot l_k &+ \left[\omega_k^2 + \left(6\xi -\frac34 + \frac{4 m^2}{ \omega_k^2} - \frac{3 m^4}{\omega_k^4}\right) \left(\frac{\dot a}{a}\right)^2 \right. \\
&\left.+ \left(6\xi - \frac32 + \frac{m^2}{\omega_k^2}\right) \frac{\ddot a}{a} \right]l_k  = 0 \ , 
\end{split}
\end{equation}
with $\xi=0$. Hence, the longitudinal polarization seems to behave, in the massless limit, as a minimally coupled scalar field. This must be checked by comparing the renormalized stress-energy tensor. 


\subsection{WKB ansatz and adiabatic expansion}

The above results for the transverse and longitudinal modes suggests the conventional WKB-type ansatz for the adiabatic expansion

\bea \label{adiabaticProca}
    h_k(t) &\sim& \frac{\exp (-i \int^t \Omega_k(t')dt')}{\sqrt{\Omega_k(t')}}, \\ l_k(t) &\sim& \frac{\exp (-i \int^t \Lambda_k(t')dt')}{\sqrt{\Lambda_k(t')}} \ , 
\eea
supplemented with 
\be 
    \mathcal {W} = \Lambda_k - \frac{i}{2}\left( \frac{\dot \Lambda_k}{\Lambda_k} +  \frac{\dot a}{a} \left(3 - \frac{2 m^2}{\omega_k^2} \right)\right) 
\ee
With this ansatz, both transverse and longitudinal modes are automatically normalized since (\ref{Wc}) (for transverse) and (\ref{normalizationReW}) (for longitudinal)   are satisfied by construction. The $\nu=0$ equation for the longitudinal polarization is also satisfied trivially, since (\ref{nuzero_longit}) holds. 
More specifically, we have
\be l_k^* \dot l_k = -\left(i + \frac{\dot \Lambda_k}{2 \Lambda_k^2}\right), \ee
and 
\bea \left.\mathfrak{Re}\mathcal{W}\right|_\text{WKB} &\sim& \frac{1}{|l_k|^2} = \Lambda_k, \\
\left. \mathfrak{Im}\mathcal{W} \right|_\text{WKB} &\sim & - \frac{1}{2}\left[ \frac{\dot \Lambda_k}{\Lambda_k} + \frac{\dot a}{a}\left(3 - \frac{2 m^2}{\omega_k^2}\right) \right],\eea
so the WKB ansatz guarantees proper normalization for the longitudinal modes. We agree with the analysis outlined in \cite{castagnino-chimento-laciana}, 
except for some signs that we interpret as misprints. 

The $\nu=j$ equations that turned into motion equations for $h_k$ and $l_k$  are (\ref{heq}) and (\ref{leq}), respectively. But these equations are identical to (\ref{waveq}) [or to Eq. (3.35) in \cite{parker-toms}] for different choices of $\sigma$.  Therefore, one can solve them adiabatically via the above WKB-type ansatz.  For our purposes, it is enough to reach the fourth adiabatic order.


\section{Two-point function and $\langle T^\alpha_\alpha \rangle$ for the Proca field}\label{regularization}

We are especially interested in evaluating $\langle T^\alpha_\alpha \rangle$. The computation of the full stress-energy tensor $\langle T_{\mu\nu} \rangle$ is doable but not necessary to obtain the trace of it. The contribution of the mass term in the Lagrangian density gives
\begin{equation}
    T_{\alpha\beta}^{(m)} = -\frac{m^2}{2} g_{\alpha\beta} A^\rho A_\rho + m^2 A_\alpha A_\beta.
\end{equation}
Hence, at quantum level we have
\be
    \langle T_\alpha^\alpha \rangle  = - m^2  g_{\alpha\beta} \langle A^\alpha A^\beta \rangle \ , 
\ee
since the other terms of the stress-energy tensor are traceless. Therefore only the two-point function is needed. 
In the renormalization process, however, it is necessary to subtract terms up to the fourth adiabatic order from the two-point function, as the divergences in $T_{\alpha \beta}$ are generically worse than in the two-point function.  
Expanding the field in modes
\be
    A^\mu =   \sum_{s=1}^3\int  d^3 k \left( A^{\mu (s)}_ {\vec k} a^{(s)}_{\vec {k}} + \text{h.c.} \right) \ , 
\ee
we get
\begin{equation}
    \langle g_{\alpha\beta}{A^\alpha A^\beta}\rangle = \sum_{s=1}^3\int  d^3 k \left(g_{\alpha\beta} A^{\alpha (s)*}_{\vec k} A^{\beta (s)}_{\vec k} \right).
\end{equation}
If one introduces the WKB ansatz and the polarization basis used above, the formal result is
\begin{equation}
    \langle  {T^\alpha_{\alpha}}\rangle = \int \frac{d^3 k}{(2\pi)^3}\left\{\frac{m^2}{a^3 \Omega_k} + \frac{\omega^2_k}{2 a^3 \Lambda_k} -\frac{ |\vec k|^2}{2 a^5 \Lambda_k \omega^2_k}|\mathcal{W}|^2 \right\}.
    \label{vector_trace_complete}
\end{equation}
From the first term it can be seen that the transverse polarizations, when taking $m\to 0$, will produce twice the trace anomaly of a conformal scalar field ($\xi=1/6$), since $\Omega_k$ obeys the equation of motion of such a field and its contribution to the two-point function is similar to that of a conformally coupled scalar.
 The longitudinal polarization is more involved, and since it does not exactly behave as  a minimally coupled scalar field, we expect some differences. Expanding up to the fourth adiabatic order, for the longitudinal polarization we have
\be \label{TLambda0-4}
\begin{split}    
    &\langle T^\alpha_\alpha \rangle^{(0-4)}_\Lambda =+\frac{7 m^2 \dot a^2}{48 \pi^2 a^2}  + \frac{m^2 \ddot a}{48 \pi^2 a}  
    - \frac{3 \dot a^4}{32 \pi^2 a^4} + \frac{13 \dot a^2 \ddot a}{480 \pi^2 a^3} \\
    &+ \frac{9 \ddot a^2}{80 \pi^2 a^2} 
    + \frac{29 \dot a \dddot a}{160 \pi^2 a^2} + \frac{3 \ddddot a}{160 \pi^2 a} \\
    &+\frac{1}{(2\pi)^2 a^3}\int_0^\infty  dk |\vec k|^2 \left[ \frac{m^2}{\omega_k} - \frac{3 m^2 \dot a ^2}{2 a^2 \omega_k^3} - \frac{m^2 \ddot a}{a \omega_k^3} \right. \\
    & \left.+\frac{\ddot a}{a \omega_k} + \frac{5 \dot a^2 \ddot a}{4 a^3 \omega^3} 
    - \frac{\ddot a^2}{4 a^2 \omega_k^3} 
    - \frac{3 \dot a \dddot a}{4 a^2 \omega_k^3} - \frac{\ddddot a}{4 a \omega_k^3}\right] \ . \ \ \ \ \ \ 
\end{split}
\ee
It is important to remark that the last momentum integral is generically divergent, even at the fourth adiabatic order. It behaves as 
\be 
 \langle T^\alpha_\alpha \rangle^{(4)}_\Lambda \sim 
\int d^3 k \left[\frac{-\Box R}{384 \pi ^3 |\vec k|^3}+\mathcal{O}(\frac{1}{k})^4 \right]
\ . \ee
This is in sharp contrast to the behavior of conformally coupled scalars or spin-$1/2$ fields \cite{LNT, RNT}. This type of divergence  caused confusion in the literature,  since  a finite behavior was ({\color{black}incorrectly}) expected in the massless limit \cite{chimento-cossarini}, in parallel to the behavior of conformally coupled scalar fields. We include in the Appendix a detailed analysis for de Sitter as a special case, since $\Box R=0$ and the fourth adiabatic order is not divergent. 


To have a proper understanding of this result,  it is  very instructive  to compare  (\ref{TLambda0-4})  with that of a minimally coupled scalar field (\ref{adbBunch}). 
In the divergent terms, the only differences arise at second adiabatic order for a few terms. 
The different divergences are multiplied by a mass, so in the massless limit the longitudinal polarization behaves exactly as a $\xi=0$ scalar. However, as the finite parts are different, these fields are not  exactly the same. If we just focus on the finite terms, we get 

\bea
    &&\text{Finite} \left( \langle T^{\alpha}_\alpha\rangle^{(0-4)}_{\Lambda}\right) - \text{Finite}\left(\langle T^{\alpha}_\alpha\rangle^{(0-4)}_{\xi=0} \right) = 
     \frac{m^2 \dot a^2}{4 \pi^2 a^2}  \nonumber   \\ &+&\frac{1}{2880\pi^2} \left[+ 60 \left(R_{\alpha\beta}R^{\alpha\beta} - \frac{1}{3}R^2\right) +20\Box R\right] \ , \eea
obtaining  a well-defined and covariant result in the massless limit. 

For the Proca field we have to distinguish the contribution of transverse modes, equivalent to two massless and conformally coupled scalar fields, and the longitudinal modes.
 The contribution of the transverse modes to the trace anomaly gives
\bea &&\langle T^{\text{Proca}}_{\text{Transverse}}
\rangle = 2\langle
T^{\text{Scalar }\xi=1/6} 
\rangle = \nonumber \\
 &&\frac{1}{2880\pi^2} \left[- 2\left(R_{\alpha\beta}R^{\alpha\beta} - \frac{1}{3}R^2\right) + 2\Box R\right]\ . \eea
Furthermore, the contribution of the longitudinal modes minus the contribution of a massless and minimally coupled field gives 
\bea 
&&\langle
T^{\text{Proca}}_{\text{Longitudinal}}
\rangle  - \langle
T^{\text{Scalar }\xi=0}
\rangle = \nonumber \\ &&\frac{1}{2880\pi^2} \left[- 60\left(R_{\alpha\beta}R^{\alpha\beta} - \frac{1}{3}R^2\right) -20\Box R\right]\ . \ \ \ \  \eea
Therefore, by combining the above results,  in the massless limit we obtain 
\bea \label{Tprocaxi}&&\langle T^{\text{Proca}}
\rangle 
  - \langle
T^{\text{Scalar }\xi=0} 
\rangle = \nonumber \\  
&&\frac{1}{2880\pi^2} \left[- 62\left(R_{\alpha\beta}R^{\alpha\beta} - \frac{1}{3}R^2\right) -18\Box R\right]\ . \ \ \ \ \eea
The right-hand side is  the usual trace anomaly of the electromagnetic field, including the  coefficient in $\Box R$.
The relation (\ref{Tprocaxi}) can also be extended to all components of the stress-energy tensor. It can be checked with the complete set of expressions derived from the adiabatic regularization method for both the minimally coupled scalar and the Proca fields. We omit the details.  Therefore, one can write, in the massless limit
\be \label{fresult}\langle T^{\text{Proca}}_{\mu\nu} \rangle = \langle T^{\text{Maxwell}}_{\mu\nu} \rangle + \langle T^{\text{Scalar }\xi=0}_{\mu\nu}\rangle \ . \ee
$\langle T^{\text{Maxwell}}_{\mu\nu} \rangle$ is univocally determined by the electromagnetic trace anomaly given in the right-hand side of (\ref{Tprocaxi}). We remark that the vacuum is  assumed to be within the family of vacuum states that obey the adiabatic  conditions (\ref{adiabaticProca}) and  (\ref{WKB}).  (For a detailed analysis of  the adiabatic condition see \cite{Nadal}.)

 \section{Comparison with other approaches}\label{comparison}

 In the St\"uckelberg procedure the action of the Proca field (\ref{Sproca}) is changed to the new action 
 \be \label{SprocaS} S_s= \int d^4x \sqrt{-g}\left[-\frac{1}{4}F_{\mu\nu}F^{\mu\nu} +\frac{1}{2}m^2 \left(A^\mu - \frac{1}{m} \partial^\mu \phi\right)^2 \right] \ . \ee 
 $S_s$ is gauge invariant under the local gauge transformations $A_\mu \to A_\mu + \partial_\mu \lambda(x)$ and $\phi \to \phi + m\lambda (x)$. In the simple gauge fixing $\phi=0$ one easily recovers the original Proca action.
 The gauge invariance can also be treated by the gauge fixing condition $\nabla_\mu A^\mu- m \phi=0$ and the gauge fixing term
 \be S_{gf}= -\frac{1}{2} \int d^4x \sqrt{-g} (\nabla_\mu A^\mu- m \phi)^2 \ . \ee 
 Adding $S_{gf}$ to (\ref{SprocaS}) transforms the action into 
 \bea \label{SprocaS2} S_s + S_{gf}&=&\frac{1}{2} \int d^4x \sqrt{-g}[A^\nu( \delta^\mu_\nu \Box + R^\mu_\nu -m^2 \delta^\mu_\nu) A_\mu \nonumber \\
 &+& \phi(\Box - m^2)\phi ]\ . \eea 
 The one-loop effective action $\bar \Gamma^{(1)}$ is obtained from the second-order operators in (\ref{SprocaS2}) together with the contribution  ghost fields. It reads \cite{buchbinder-shapiro21, buchbinder-berredo-shapiro}
 \bea \bar \Gamma^{(1)} &=& \frac{i}{2} \mathrm{Tr}  \log ( \delta^\mu_\nu \Box + R^\mu_\nu -m^2 \delta^\mu_\nu) +\frac{i}{2} \mathrm{Tr} \log (\Box - m^2) \nonumber \\
 &-& i\mathrm{Tr} \log (\Box -m^2)\ , \eea 
 where the last term is the contribution of the two ghost scalars (with the opposite sign due to their  odd Grassmann   statistics). Furthermore, the first and third terms combine, in the massless limit $m\to 0$, to give the effective action of the massless Maxwell field
\bea \bar \Gamma^{(1)} _{\text{Maxwell}} &=& \frac{i}{2} \mathrm{Tr}  \log ( \delta^\mu_\nu \Box + R^\mu_\nu) - i\mathrm{Tr} \log \Box \ . \eea 
 The second term gives, in this limit,  the contribution of the massless scalar field. Therefore, in the massless limit, we get
  \be \bar \Gamma^{(1)}_{\text{Proca}} = \bar \Gamma^{(1)}_{\text{Maxwell}} + \bar \Gamma^{(1)}_{\text{Scalar } \xi=0}  \ . \ee
 This is the analog, in the language of the effective action, of our results (\ref{Tprocaxi}) and (\ref{fresult}).
 
\subsection{De Sitter space}

As a second test of  the adiabatic method,  it is interesting to look at its particularization in de Sitter space with $a(t)=e^{Ht}$. The first important point to consider is that the ultraviolet divergences  cancel out exactly at the fourth adiabatic order, since $\Box R = 0$. Therefore, both $ \langle T^{\text{Proca}}\rangle$ and $\langle
T^{\text{Scalar }\xi=0}\rangle $ are, by itself, finite. This is what happens generically, for arbitrary expansion factor $a(t)$, for conformally coupled scalars and spin-$1/2$ fields.

For the Proca field we found the following finite contributions for the subtraction terms at fourth adiabatic order 
\be \langle T^\alpha_\alpha \rangle^{(4)}_\Lambda = \frac{59 H^4}{240\pi^2}  \ee
\begin{equation}
    \langle T^\alpha_{\alpha}\rangle^{(4)}_\Omega=  -\frac{H^4}{120 \pi^2} 
\end{equation}
These yield to the following trace  in the massless limit  
 \bea \label{Procadesitter}\langle T^\alpha_\alpha \rangle &=& -\langle T^\alpha_\alpha \rangle^{(4)}_\Lambda- \langle T^\alpha_{\alpha}\rangle^{(4)}_\Omega 
 = -\frac{19 H^2}{80\pi^2} \ . 
 \eea
Using now the general result    (\ref{Tprocaxi})  we predict
  \bea \label{ABresult} \langle T^{\text{Scalar }\xi=0} 
\rangle = -\frac{119H^4}{240 \pi^2}  \ . 
 \eea
A direct calculation shows the fourth adiabatic order of the stress-energy tensor for a Proca theory, in the massless limit, to be proportional to the metric. Since that has to also be the case for the Maxwell theory, due to the conformal nature of the fields and the de Sitter symmetry group, we can conclude, using (\ref{Tprocaxi}), that 
\bea \label{ABresult2} \langle T^{\text{Scalar }\xi=0}_{\mu\nu} 
\rangle = -\frac{119H^4}{960\pi^2} g_{\mu\nu} \ . 
 \eea

The result (\ref{ABresult2}) 
is in exact agreement with the vacuum expectation values of the stress-energy tensor evaluated for the Hadamard vacua $\ket{0}_{A,B}$ proposed in \cite{allen-folacci}. 
As the above prediction only considers the adiabatic subtractions but yields the correct result, we conclude that our result is compatible with noncontribution to the anomaly from the modes. This was tested numerically with the exact mode functions, reaching the same conclusion. We include the details in the Appendix.

\section{Summary and conclusions}\label{conclusions}

The renormalization mechanism usually requires one to identify first the ultraviolet divergences and then proceed to subtract them consistently. For  cosmological FLRW spacetimes, adiabatic regularization is likely the most direct and intuitive method. When it was first introduced, it was not known that the conformal anomaly existed. The conformal anomaly for scalar fields was first derived within the adiabatic method in \cite{Hu}. Since  the adiabatic regularization method requires one to start with a nonvanishing mass, the conformal anomaly was  obtained  by taking the massless limit at the end of the calculations. The finite remaining integrals of the fourth adiabatic order for the stress-energy tensor lead to the expression of the trace anomaly. This strategy also works for spin-$1/2$ fields \cite{LNT}. However, for spin-$1$ fields the problem is more tricky, since adding a mass to the field changes the number of physical degrees of freedom, transforming  a gauge invariant field into a Proca field. Furthermore, as is known from the form of the two-point function in flat space, the massive theory becomes more singular. 
In a FLRW spacetime, the fourth-order adiabatic terms of the stress-energy tensor are ultraviolet divergent for a massive spin-$1$ field, in sharp contrast to the finite integrals obtained for conformally coupled spin-$0$ \cite{Hu} and spin-$1/2$ \cite{LNT}.     
Therefore, obtaining the trace anomaly for spin-$1$ fields by the usual rule (i.e., setting the mass at the end of the calculation) is inconsistent. A way out of this tension is to assume a massless field, implement gauge fixing, and add ghosts. Then artificially add masses to the gauge field and ghosts, then do the adiabatic expansion and set the masses to zero at the end of the calculation. This is the strategy used in \cite{chimento-cossarini, chu-koyama}. This makes the whole procedure  quite complicated and deviates from the simplicity of the original adiabatic method. 

In this paper we have followed a different path, closer to the original philosophy of adiabatic regularization. 
The divergences present in the fourth-order adiabatic terms of the stress-energy tensor arise from the longitudinal polarization of the Proca field. The important observation is that these divergences coincide exactly with similar ultraviolet divergences found for a minimally coupled scalar field (when the mass is set to zero at the end of the calculation). {\color{black} As a consequence, subtracting the quantities $\langle T^{\text{Proca}}_{\mu\nu} \rangle -\langle T^{\text{Scalar }\xi=0}_{\mu\nu}\rangle$
and setting at the end $m=0$ we exactly get the renormalized stress-energy tensor of the Maxwell field and the corresponding trace anomaly, as it has been anticipated in (\ref{main1}).}  We consider this to be the simplest
adiabatic regularization description of a Proca field and fills a gap in the existing literature on the adiabatic method.
It is  our main result, and it is consistent with effective action methods \cite{buchbinder-shapiro21, buchbinder-berredo-shapiro}, as shown in Sec. \ref{comparison}. 
We have also checked that our result also fits  well with  well-founded results in the literature on field  quantization in de Sitter space \cite{allen-folacci}.

\section*{Acknowledgments}

We thank A. Ferreiro for collaboration in the early stages  of this project. This work is supported by the Spanish Grants No. PID2020-116567GBC2-1  and No. PROMETEO/2020/079 (Generalitat Valenciana). 
Some of the computations have been done with the help of \textit{Mathematica}\texttrademark.

\appendix

\section{Some calculations in de Sitter space for a Proca field} \label{dsitter}

In this appendix we give some details on the calculation of the renormalized trace $\langle T^\alpha_\alpha \rangle $  for a special adiabatic vacuum in de Sitter,  analogous to the Bunch-Davies one for massive scalar fields \cite{Bunch-Davies}. For the analysis of the modes in the helicity-momentum basis we will follow \cite{cotaescu} with the conformal form of the metric 
\begin{equation}
    ds^2 = \frac{1}{(H t_c)^2}(dt_c^2 - d \vec{x}^2).
\end{equation}
The contribution of the modes to the two-point function is given by
\bea g^{\mu\nu}\langle A_\mu A_\nu \rangle &=& (H t_c)^2 \int d^3 k \{ |\gamma(t_c)_k|^2 - |\beta(t_c)_k|^2 \nonumber \\
    &-& 2|\alpha(t_c)_k|^2 \} \ , \eea
where these functions satisfy
\begin{equation}
\begin{split}
    & \ddot \alpha_k + \left( |\vec k|^2 + \frac{m^2}{H^2 t_c^2} \right) \alpha_k = 0 \\
    & \beta_k = -\frac{i}{k}\left( \dot \gamma_k - \frac{2}{t_c} \gamma_k \right) \\
    & t_c^2 \ddot \gamma_k - 2 t_c \dot \gamma_k + \left( t_c^2 |\vec k|^2 + \frac{m^2}{H^2} + 2\right) \gamma_k = 0 \ .
\end{split}
\end{equation}
The equation for $\gamma_k$ can be transformed into a Bessel by a simultaneous rescaling $\gamma_k = (-t_c)^{3/2} G_k$ and change of variables $\eta = - k t_c$. The procedure for the $\alpha_k$ equation is identical, but with exponent 1/2. Then we obtain
\begin{equation}
    \eta^2 \frac{d^2 G_k}{d\eta^2} + \eta \frac{d G_k}{d\eta} + \left(\eta^2 - \nu^2 \right) G_k, \;\;\; \nu=\sqrt{\frac{1}{4}-\frac{m^2}{H^2}}.
\end{equation}
Since we are interested in the limit $m\to 0$, then we take $1/2 \geq \nu \geq 0$. The solutions, with the appropriate asymptotic behavior for vacuum modes at $t_c \to -\infty$, are
\begin{equation}
\begin{split}
    \alpha(t_c)_k &= N_1 e^{i\frac{\pi \nu}{2} } (-t_c)^{1/2} H^{(1)}_\nu(-k t_c) \\
    \beta(t_c)_k &= i N_2 e^{i\frac{\pi \nu}{2} } \left\{ \frac{1}{k} \left(\nu -\frac{1}{2}\right) (-t_c)^{1/2} H^{(1)}_\nu(-k t_c)\right. \\
    &\left.- (-t_c)^{3/2} H^{(1)}_{\nu+1}(-k t_c) \right\}\\
    \gamma(t_c)_k &= N_2 e^{i\frac{\pi \nu}{2} } (-t_c)^{3/2} H^{(1)}_\nu(-k t_c) .
\end{split}
\end{equation}
The normalization constants are
\begin{equation}
    N_1 = \frac{\sqrt{\pi}}{2 \sqrt{(2 \pi)^3}}, \;\; N_2 = N_1 \frac{H k}{m}.
\end{equation}
Since we expect to recover twice the scalar trace anomaly from the transverse polarizations, we focus for now only in the longitudinal one. Its renormalized contribution to the trace is given by  
\begin{equation}
\begin{split}    
   \bra{0} T^\alpha_\alpha\ket{0}_\text{$\Lambda$ {\color{red}}} & = - m^2 (H t_c)^2 \int d^3 k \left\{ |\gamma(t_c)_k|^2 - |\beta(t_c)_k|^2 \right\} \\
    & - \left( \braket{T^\alpha_\alpha}_\Lambda^{(0)} + \braket{T^\alpha_\alpha}_\Lambda^{(2)}+ \braket{T^\alpha_\alpha}_\Lambda^{(4)} \right),
\end{split}
\end{equation}
where the fourth adiabatic order is convergent for de Sitter. Consider for now the longitudinal mode contribution. After changing variables to $\eta = -k t_c$,
\begin{equation}
\begin{split}    
    \braket{T^\alpha_\alpha}_\Lambda = - \frac{H^4}{8 \pi} & \int_0^\infty \eta^4 d \eta  \left\{ \left[ 1- \left( \frac{\nu - 1/2}{\eta} \right)^2 \right]|H_\nu^{(1)}(\eta)|^2 \right. \\ 
    &  - |H_{\nu+1}^{(1)}(\eta)|^2 \\
    & \left. + \left(\frac{\nu - 1/2}{\eta}\right) 2 \mathfrak{Re}\left[ H_\nu^{(1)}(\eta)^* H_{\nu+1}^{(1)}(\eta) \right] \right\}.
\end{split}
\end{equation}
The contribution from the transverse modes is significantly simpler
\begin{equation}
    \braket{T^\alpha_\alpha}_\Omega=\frac{H^4}{4\pi} \int^\infty_0 \left(\frac{m^2}{H^2}\right) \eta^2 d \eta |H^{(1)}_\nu|^2.
\end{equation}

We want to check if there is a contribution to the renormalized trace arising from the modes or lower adiabatic order. A numerical approach is sufficient to test that. First, we express the adiabatic subtractions using $\eta=-k t_c$ and $z=m^2/H^2$,
\begin{equation}
\begin{split}
    &\braket{T^\alpha_\alpha}_\Lambda^{(0-4)} =  \braket{T_\alpha^\alpha}^{(0)}_{\Lambda}+\braket{T_\alpha^\alpha}^{(2)}_{\Lambda}+\braket{T_\alpha^\alpha}^{(4)}_{\Lambda}  \\
    &= H^4 \int^\infty_0 \frac{\eta^2 d \eta}{512 \pi ^2 \left( \eta ^2+z\right)^{13/2}} \\
    &\times\left\{16 \left(2 \eta ^2+z\right) \left(4 \eta ^4+z^2\right) \left(\eta ^2+z\right)^3 \right. \\
    & \;+z \left(1440 \eta ^8-340 \eta ^4 z^2+28 \eta
   ^2 z^3+3 z^4-80 \eta ^6 z\right) \\
   &\left.+128 z \left(\eta ^2+z\right)^6\right\},
\end{split}
\end{equation}
\begin{equation}
\begin{split}
    &\braket{T^\alpha_\alpha}_\Omega^{(0-4)} =\braket{T^\alpha_\alpha}^{(0)}_{\Omega}+\braket{T_\alpha^\alpha}^{(2)}_{\Omega}+\braket{T_\alpha^\alpha}^{(4)}_{\Omega} \\
    &=  H^4 \int^\infty_0 \frac{z \eta^2 d \eta}{256 \pi ^2 \left(\eta ^2+z\right)^{13/2}} \\
    &\times\left\{3 z \left(-160 \eta ^6-4 \eta ^2 z^2+z^3+220 \eta ^4 z\right) \right. \\
    & \;\left.+128 \left(\eta ^2+z\right)^6+16 z \left(6 \eta ^2+z\right)
   \left(\eta ^2+z\right)^3 \right\}.
\end{split}
\end{equation}
Then, we can express the renormalized trace as the contribution from the modes minus the adiabatic subtractions
\begin{equation}
\begin{split}    
    \braket{T^\text{Proca}} &= \braket{T^\alpha_\alpha}_\Lambda + \braket{T^\alpha_\alpha}_\Omega  - \left( \braket{T^\alpha_\alpha}_\Lambda^{(0-4)} + \braket{T^\alpha_\alpha}_\Omega^{(0-4)} \right) 
\end{split}
\end{equation}

We solve the integral numerically, for a small but nonzero value of $z$. Starting from $z=10^{-1}$, we keep reducing it by factors of 10 until we obtain the values given below, for $z=10^{-9}$. The integrand falls rapidly to zero, for values $\eta \sim 10$ in the longitudinal and transverse cases. However, due to numerical instabilities, it will blow up to infinity if $\eta$ is allowed to be big enough, since the oscillatory behavior of the integrand is not suppressed to all values of $\eta$ in the numerical approach. We know this is an artifact of the numerical method because the UV behavior of the integrand has been checked analytically, and it is like $O(1/\eta)^2 \,d\eta $ in both cases. Denoting the upper limit of the integral as $u$, and the integrand at that point as $\text{intg}(u)$, the values we used were $z=10^{-9}$, $u=40$, $\text{intg}(u)=10^{-20}$ for the transverse; and $z=10^{-9}$, $u=8$, $\text{intg}(u)=10^{-10}$ for the longitudinal. The renormalized trace for the massless Proca theory is
\begin{equation}
    \braket{T^\text{Proca}} = \frac{H^4}{240 \pi^2} (+2 -59) = -\frac{57 H^4}{240 \pi^2} \ ,
\end{equation}
where the numerical results were rounded from $2.0000012 \approx 2$ and $-58.999914 \approx -59$. Therefore, the result of the numerical calculation concurs with the prediction (\ref{Procadesitter}), with no contribution from the mode functions. Recalling that 
\begin{equation}
\begin{split}
    \braket{T^{\text{Maxwell }}} &= + \frac{62 H^4}{240 \pi^2} \ ,
\end{split}
\end{equation}
we obtain (\ref{ABresult}) to guarantee 
\begin{equation}
    \braket{T^\text{Proca}}-\braket{T^{\text{Scalar } \xi=0}} = \braket{T^{\text{Maxwell }}} \ ,
\end{equation}
the main result of the article for the trace when $m\to 0$.

\end{document}